\documentclass[twocolumn,preprintnumbers,amsmath,amssymb]{revtex4}
\usepackage{graphicx}
\usepackage{epsfig}
\usepackage{amsmath}
\usepackage{amssymb}
\usepackage{dcolumn}

\begin{document}

\title{First Order Calculation of the Inclusive Cross Section $pp\rightarrow ZZ$ by Graviton Exchange in Large Extra Dimensions}

\author{Martin Kober${}^1$}
 \email{kober@th.physik.uni-frankfurt.de}
\author{Benjamin Koch${}^{1,2}$}
\author{Marcus Bleicher${}^1$}
 
\affiliation{${}^1$~Institut f\"ur Theoretische Physik, Johann Wolfgang Goethe-Universit\"at, 
Max-von-Laue-Str.~1, 60438 Frankfurt am Main, Germany \\
${}^2$~Frankfurt Institute for Advanced Studies (FIAS), Max-von-Laue-Str.~1, 
60438~Frankfurt am Main, Germany}
\date{\today}

\begin{abstract}
We calculate the inclusive cross section of double Z-boson production within large extra dimensions at
the Large Hadron Collider (LHC). Using perturbatively quantized gravity in the ADD model we
perform a first order calculation of the graviton mediated contribution to the $pp\rightarrow
ZZ+x$ cross section. At low energies (e.g. Tevatron) this additional contribution is very
small, making it virtually unobservable, for a fundamental mass scale above 
$2500$~GeV. At LHC energies however, the calculation indicates that the ZZ-production rate within the ADD model
should differ significantly from the Standard Model if the new fundamental mass
scale would be below $15000$~GeV. A comparison with the observed production rate at the LHC 
might therefore provide direct hints on the number and structure of the extra dimensions.
\newline
\end{abstract}

\maketitle

\section{Introduction}

The possible existence of additional spatial dimensions has been a fascinating topic for
theoretical physicists since the early ideas of Kaluza and Klein
\cite{Kaluza:1921,Klein:1926}. The explanation why the additional dimensions have not been
discovered so far can be given by the assumption that the additional dimensions are
compactified to a very small radius. Usually the length scale of the compactified dimensions
is assumed to be of the order of the Planck scale $M_P\sim 10^{19}$~GeV and therefore far
away from the scope of experimental physics.

However there is the possibility that the additional dimensions could be realized in a different way
\cite{Antoniadis:1997zg,Antoniadis:1998ig,Arkani-Hamed:1998rs,Randall:1999ee,Randall:1999vf}. In the ADD model \cite{Arkani-Hamed:1998nn} the assumption is that the additional dimensions could be realized with a much larger
compactification radius and much smaller fundamental scales, thus opening the possibility to
observe deviations from the Standard Model (SM) at energies available at the LHC. According
to the ADD model all matter fields of the SM live on the (3+1)-dimensional submanifold
describing usual space-time and only gravity can propagate into the extra dimensions. 
This model yields an explanation why gravity appears to be so weak compared to the other
interactions existing in nature (thus circumventing the hierarchy problem). The ADD model
goes hand in hand with the definition of a higher dimensional Planck mass $M_D$ related to
the four dimensional Planck mass $M_P$ by the following equation

\begin{equation}
M_P^2=8 \pi M_D^{d+2} R^d\quad,
\end{equation}
where $d$ is the number of extra spatial dimensions and $R$ is their size. This effective reduction 
\footnote{To be specific, it is not the lowering of the Planck scale, but the
increase of the available phase space that increases the effective coupling to gravitons
within the ADD model. At high enough energies, this is equivalent to a reduction of the
Planck scale.} 
of the Planck scale leads to interesting effects for the production of
gravitons. While in usual perturbatively quantized gravity the effects of incorporating
virtual gravitons can be neglected because they are suppressed by factors of $1/M_P$ the
situation with the ADD model is different. By introducing additional dimensions according to
the ADD model, the production and exchange of gravitons can lead to observable effects \cite{Giudice:1998ck,Giudice:2000av,Han:1998sg,Balazs:1999ge,
Hewett:1998sn,Hewett:2002hv,Mirabelli:1998rt,Rubakov:2001kp,Cullen:1999hc,Cullen:2000ef,Banks:1999gd,Hall:1999mk,Barger:1999jf,Rizzo:1998fm,Emparan:2001kf,Agashe:1999qp}.

The most prominent prediction of the ADD model is probably the possible production of
microscopical black holes in upcoming collider experiments
\cite{Banks:1999gd,Giddings:2001bu,Dimopoulos:2001hw,Eardley:2002re,Bleicher:2001kh,Hossenfelder:2001dn}. As
exciting the production (and subsequent evaporation) of a black hole in the laboratory might
be, as a non-perturbative process in quantum gravity it is difficult to make quantitative
predictions for the measurement of such events. The sum of these difficulties makes most
predictions which are derived from black hole production model dependent. It is therefore
desirable to have complementary observables which allow to test a specific model in a more
quantitative way.

One possible approach that allows for a perturbative calculation is to consider the much
larger number of possible couplings between the Kaluza-Klein towers in the perturbative
gravitation sector of the ADD model and the SM. This results for instance in an enhanced
gravitational radiation into the extra dimensions \cite{Galtsov:2001iv,Cardoso:2002pa,Koch:2005bc,Ruppert:2005st,Koch:2006yi}
becoming more and more important at higher energies. It also leads to new contributions to SM
physics due to virtual graviton exchange \cite{Giudice:1998ck,Han:1998sg,Hewett:2002hv}. Such
contributions are mostly undetectable because of the much larger SM background. Ideally one
is looking for SM processes that have a very clear experimental signal but a very low cross
section. Further, those processes should also have some tree-level contributions in the ADD
model.

In this paper we suggest to look for the production of Z-boson pairs,
as it is done for a stabilized Randall-Sundrum scenario in \cite{Park:2001vk}. 
For the case of the ADD model in \cite{Balazs:1999ge} there is already considered production of a Z-pair 
by graviton exchange for the LEP and the Tevatron on the one hand and the decay mode $Z \rightarrow f \bar f+\mathcal{G}$ 
on the other hand. According to these considerations there are obtained constraints for $M_D$ below 1 TeV.
However, we show that ZZ-production at the LHC can lead to constraints for $M_D$ far above 1 TeV.

In the SM the process of ZZ-production is suppressed due to the two electroweak couplings and due to the large
ZZ-production threshold. Another nice feature of this process at lowest order in the
perturbative expansion is that the ADD contribution completely decouples from the SM cross
section and can be treated separately. 

In the following section we present the prerequisites for the effectively quantized gravity
calculation in higher dimensions. After this we will consider ZZ-production in (anti-)proton-proton
collisions by graviton exchange at tree-level. We will see that at least in case of a D-dimensional Planck mass 
$M_D$ in the ~TeV region the extra dimensional contribution to the SM cross section leads to a substantial deviation from the
well known ZZ-production rate of the SM.

\section{Effective quantum field theory of gravity in higher dimensions}

Let us shortly remind the reader on the effective field theory of gravity in extra dimensions
given in \cite{Giudice:1998ck,Han:1998sg,Callin:2004zm}. Here, a small perturbation $h_{MN}$ of the metric of flat
Minkowski space-time $\eta_{MN}$ is considered. The complete metric $g_{MN}$ is then given by 

\begin{equation}
g_{MN}=\eta_{MN}+2M_D^\frac{(D-2)}{2} h_{MN}\quad.
\label{Entwicklung-Metrik}
\end{equation}
In the present case of additional dimensions the indices $M$ and $N$ run from 1 to $D=(3+d)+1$.
According to this a space-time point is described by a D-tupel $(t,{\bf x},y_4,...,y_d)$. 
The Einstein-Hilbert action according to the Einstein equations in D-dimensions reads

\begin{equation}
S=\frac{1}{2}M^{D-2} \int d^D x \sqrt{-g} R\quad,
\label{EinsteinHilbert}
\end{equation}
where $g$ describes the determinant of the metric.
Using the expansion (\ref{Entwicklung-Metrik}) in (\ref{EinsteinHilbert}) leads to the following Lagrangian

\begin{eqnarray}
\mathcal{L}_h=-\frac{1}{2}\partial_M \partial^M h+\frac{1}{2}\partial_R h_{MN} \partial^R h^{MN}\nonumber\\
+\partial_M h^{MN}\partial_N h-\partial_M h^{MN} \partial_R h^R_N\quad.
\end{eqnarray}
The $d$ additional dimensions are assumed to be compactified to a $d$-dimensional torus with
radii $R$ meaning that the coordinates belonging to the additional dimensions are periodic
with respect to the transformation $y_j \rightarrow y_j+2\pi R, j=4,...,d$. Thus, the
perturbation of the metric field can be expressed as

\begin{equation}
h_{MN}(z)=\sum_{j=1}^d \sum_{n_d=-\infty}^{\infty} \frac{h_{MN}(x)}{\sqrt{V_d}}e^{i\frac{n^j y_j}{R}}\quad.
\end{equation}
This leads to an effective Lagrangian on the 3+1-dimensional submanifold of the following form

\begin{eqnarray}
\mathcal{L}_h=-\frac{1}{2}\partial_\mu \partial^\mu h+\frac{1}{2}\partial_\rho h_{\mu\nu} \partial^\rho h^{\mu\nu}
+\partial_\mu h^{\mu\nu}\partial_\nu h\nonumber\\
-\partial_\mu h^{\mu\nu} \partial_\rho h^\rho_\nu-\frac{1}{2} m^2 (h^{\mu\nu}h_{\mu\nu}-h^2)\quad.
\end{eqnarray}
The occurring mass of the graviton corresponds to the excitations of the gravitational field
in the compactified dimensions and is related to them according to the equation

\begin{equation}
m_n^2=\sum_{j=1}^{d}\left(\frac{n_j^2}{R^2}\right)\quad.
\end{equation}
By following the path integral quantization procedure one is lead to the expression for the graviton propagator. The effective mass of the graviton breaks the usual gauge invariance of gravitation

\begin{equation}
x_\mu \rightarrow x_\mu+\epsilon_\mu \quad,\quad h_{\mu\nu}\rightarrow
h_{\mu\nu}-(\partial_\mu \epsilon_\nu+\partial_\nu \epsilon_\mu)\quad,
\end{equation}
and thus the Faddeev-Popov procedure can be omitted. The graviton propagator is then

\begin{equation}
\Delta_{\mu\nu\rho\sigma}(x,y)=\int \frac{d^4 k}{(2\pi)^4}\frac{P_{\mu\nu\rho\sigma}(k)}{k^2-m_n^2}e^{-ik(x-y)},
\label{GravitonpropagatorMasse}
\end{equation}
with polarization-tensor

\begin{eqnarray}
P_{\mu\nu\rho\sigma}(k)=\frac{1}{2}(\eta_{\mu\rho}\eta_{\nu\sigma}+\eta_{\mu\sigma}\eta_{\nu\rho})
-\frac{1}{2}\eta_{\mu\nu}\eta_{\rho\sigma}\\\nonumber
-\frac{1}{2m^2}(\eta_{\mu\rho}k_\nu k_\sigma+\eta_{\mu\sigma}k_\nu k_\rho+\eta_{\nu\rho}k_\mu k_\sigma+\eta_{\nu\sigma}k_\mu k_\rho)\\\nonumber
+\frac{1}{6}(\eta_{\mu\nu}+\frac{2}{m^2} k_\mu k_\nu)(\eta_{\rho\sigma}+\frac{2}{m^2}k_\rho k_\sigma)\quad.
\label{Polarisationstensorgraviton}
\end{eqnarray}
By addition of the terms arising from the energy-momentum tensor of the matter and
interaction fields \cite{Giudice:1998ck} the vertices are obtained.  In the limit of a weak
gravitational field only the matter and interaction fields of the SM fields (assumed to live on the 3+1-dimensional submanifold) contribute to the energy momentum tensor. It is therefore given by
\begin{equation}
T_{AB}(z)=\eta^\mu_A \eta^\nu_B T_{\mu\nu}(x)\delta(y)\quad.
\end{equation}

\section{S-matrix and cross section}

In the perturbative approach we use the SM couplings $\sqrt{\alpha_{ew,strong}}$ and the
ratio $m_{X}/M_D$ as smallness parameters, where $m_{X}$ stands for the mass scale in the
process (in our case essentially the Z-boson). First, the collision processes of the partons, the
quarks and gluons within the proton, have to be regarded. We restrict our analysis to
processes at tree-level.

\begin{figure}[ht]
\centering
\epsfig{figure=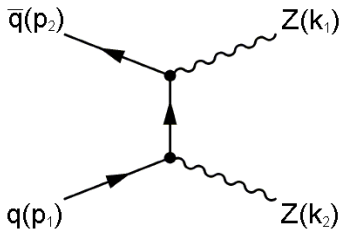,width=4.5cm}
\caption{\label{F2} Standard model contribution due to a t-channel quark exchange.
The quark line $q$ can represent any Standard Model quark.}
\end{figure}

\begin{figure}[ht]
\centering
\epsfig{figure=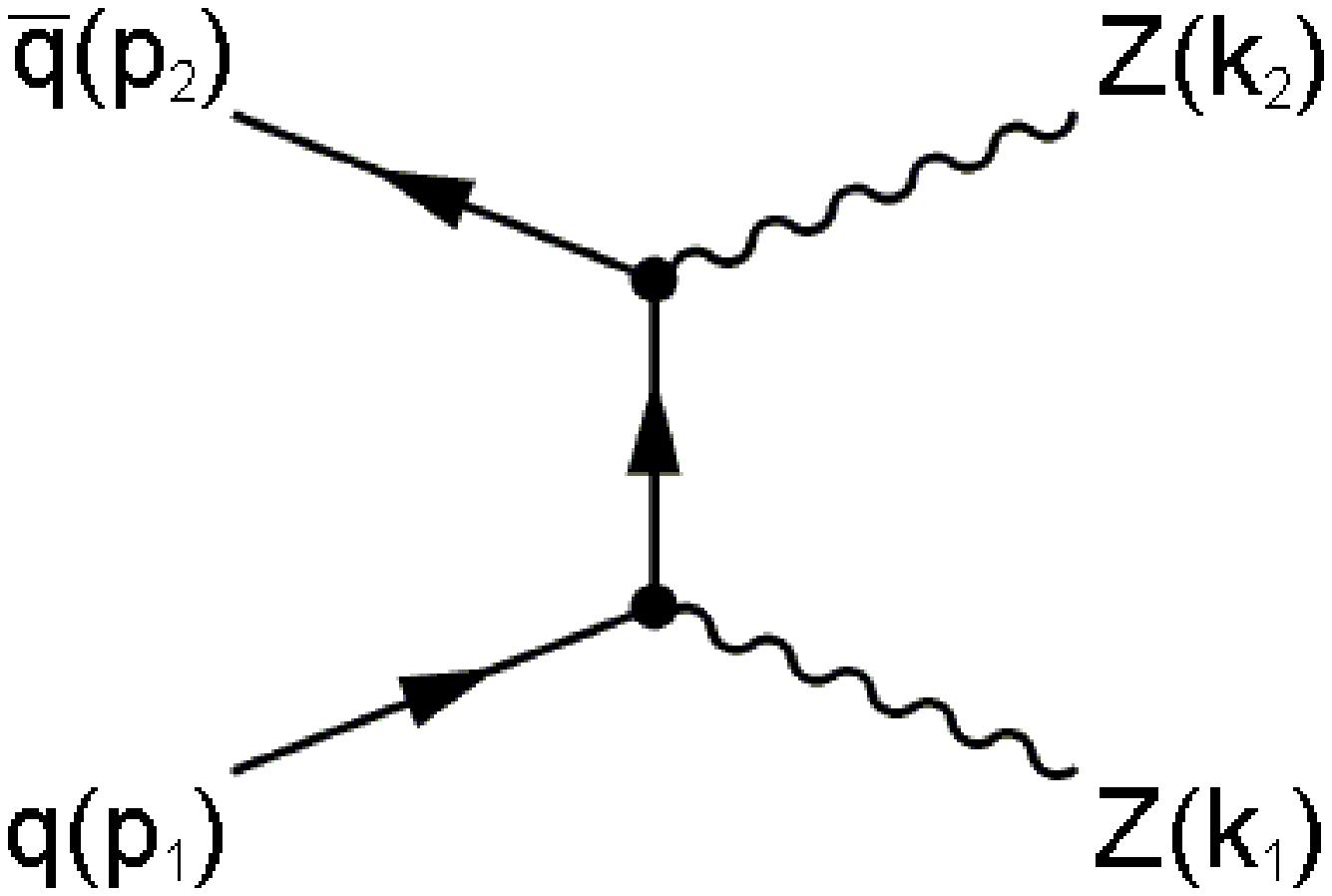,width=4.5cm}
\caption{\label{F3} Standard model contribution due to a t-channel quark exchange and
crossed final momenta $k_1$ and $k_2$.
The quark line $q$ can represent any Standard Model quark.}
\end{figure}

\begin{figure}[ht]
\centering
\epsfig{figure=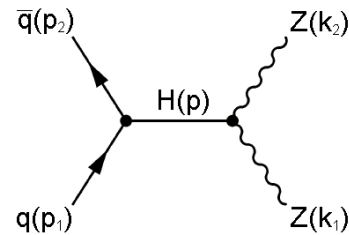,width=4.5cm}
\caption{\label{F4} Standard Model contribution due to Higgs boson in the s-channel.
The quark line $q$ can represent any Standard Model quark.}
\end{figure}

The processes corresponding to the Feynman diagrams displayed in figures (\ref{F2}),
(\ref{F3}) and (\ref{F4}) contribute to the cross section for the quark anti-quark
process of the SM. In leading order, there is no contribution from the gluons in the SM. Although
SM calculations to higher orders in perturbation theory are feasible \cite{Baur:2001ze},
they are neglected here, because one would also have to do higher orders in the ADD
extension for consistency. However, as the ADD extension of the SM is non-renormalizable,
those higher order calculations would not provide new insight or better accuracy. We will
therefore limit ourselves to tree-level calculations, which are standard and will not be
shown in detail here. As the parton distribution functions are defined in the high energy
limit for massless quarks, we finally take quark masses to be zero.

\begin{figure}[ht]
\centering
\epsfig{figure=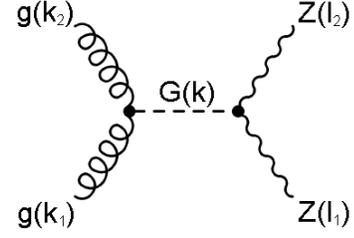,width=4.5cm}
\caption{\label{F1} ZZ-production due to the annihilation of two gluons
into a virtual graviton}
\end{figure}

Concerning the contributions through graviton mediation all processes induced by a quark anti-quark 
pair can be neglected.
The probability for the transition of a quark anti-quark pair to a ZZ-pair mediated by a single
graviton $S_{graviton}(q+\bar q \rightarrow ZZ)$ is exactly zero
and all other processes leading to such a transition and involving a graviton are of higher order.  
Therefore, only the process where gluons annihilate and produce a graviton which couples
to two Z-bosons remains at this order of perturbative expansion. 
This process has different external particles than the SM process 
where only the quarks give a contribution to the double Z-boson production rate. 
Thus, the contribution by graviton exchange can be regarded separately. 
At tree-level, the S-Matrix for the transition from a
gluon pair to a ZZ-pair by mediation of a graviton corresponding to the Feynman graph 
(Figure (\ref{F1})) is given by 

\begin{eqnarray}
S_{graviton}&(g(k_1,g_1)+g(k_2,g_2) \rightarrow Z(l_1,Z_1)+Z(l_2,Z_2))\nonumber\\
=&\frac{1}{(2\pi)^4}\int d^{4}k \frac{g_{1\alpha a}}{(2\pi)^\frac{3}{2}\sqrt{k_{10}}} \frac{g_{2\beta b}}{(2\pi)^\frac{3}{2}\sqrt{k_{20}}}\nonumber\\&
\left(-\frac{i}{\bar M_P}\delta^{ab}\left[W^{\mu\nu\alpha\beta}+W^{\nu\mu\alpha\beta}\right] \right)\nonumber\\&
\cdot (2\pi)^4\delta^4(k_1+k_2-k)\nonumber\\&
\sum_n \frac{iP_{\mu\nu\rho\sigma}}{k^2-m_n^2}
(2\pi)^4\delta^4(k-l_1-l_2)\nonumber\\&
\cdot \left(-\frac{i}{\bar M_P}\left[W^{\rho\sigma\gamma\delta}+W^{\sigma\rho\gamma\delta}\right] \right)\nonumber\\&
\frac{Z_{1\gamma}}{(2\pi)^\frac{3}{2}\sqrt{l_{10}}} 
\frac{Z_{2\delta}}{(2\pi)^\frac{3}{2}\sqrt{l_{20}}}\quad,
\label{S-Matrix}
\end{eqnarray}
where $\bar M_P=M_P/\sqrt{8\pi}$ and $W^{\mu\nu\alpha\beta}$ 
is defined as
\begin{eqnarray}
W^{\mu\nu\alpha\beta}\ \ \ \ \ \ \ \ \ \ \ \ \ \ \ \ \ \ \ \ \ \ \ \ \ \ \ \ \ \ \ \ \ \ \ \ \ \ \ \ \ \ \ \ \ \ \ \nonumber\\
=\frac{1}{2} \eta^{\mu\nu}(k_1^{\beta}k_2^{\alpha}-k_1\cdot k_2 \eta^{\alpha\beta})+\eta^{\alpha\beta} k_1^{\mu} k_2^{\nu}\nonumber\\
+\eta^{\mu\alpha}(k_1 \cdot k_2 \eta^{\nu\beta}-k_1^{\beta} k_2^{\nu})-\eta^{\mu\beta} k_1^{\nu} k_2^{\alpha}\quad.\nonumber\\
\label{Wexpression}
\end{eqnarray}
according to the expressions for the vertices found in \cite{Giudice:1998ck}.
Note that $g_{1\alpha a/2\beta b}$ and $Z_{1\gamma/2\delta}$ denote the 
polarization vectors of the gluons and Z-bosons respectively, where greek letters denote Lorentz-indices and latin letters 
as indices refer to the internal colour-space of QCD. 
Further $k_1$ and $k_2$ denote the initial momenta of the gluons and $Z_1$ and $Z_2$ denote the final momenta of the Z-bosons. 
The expression for the S-matrix (\ref{S-Matrix}) can be transformed to

\begin{eqnarray}
S_{graviton}&=&\frac{-i}{(2\pi)^2 \sqrt{2k_{10}} \sqrt{2k_{20}} \sqrt{2l_{10}} \sqrt{2l_{20}}} \nonumber\\ 
&&\cdot \frac{1}{\bar M_P^2} \sum_n \frac{1}{p^2-m_n^2} g_{1\alpha} g_{2\beta} \left[W^{\mu\nu\alpha\beta}+W^{\nu\mu\alpha\beta}\right] \nonumber\\
&&\cdot P_{\mu\nu\rho\sigma} \left[W^{\rho\sigma\gamma\delta}+W^{\sigma\rho\gamma\delta}\right] Z_{1\gamma} Z_{2\delta} \nonumber\\
&&\cdot \delta^4(p-l_1-l_2)\quad,
\label{S-Matrix2}
\end{eqnarray}
where we have integrated over the first delta-function and p defined as $p=k_1+k_2$.
Using the symmetry of $P_{\mu\nu\rho\sigma}$ in $\mu$,$\nu$ and $\rho$,$\sigma$ respectively, the expression (\ref{S-Matrix2}) reads

\begin{eqnarray}
S_{graviton}&=&\frac{-4i}{(2\pi)^2 \sqrt{2k_{10}} \sqrt{2k_{20}} \sqrt{2l_{10}} \sqrt{2l_{20}}} \frac{1}{\bar M_P^2} \sum_n \frac{1}{p^2-m_n^2}\nonumber\\ 
&&g_{1\alpha} g_{2\beta} W^{\mu\nu\alpha\beta} P_{\mu\nu\rho\sigma} W^{\rho\sigma\gamma\delta} Z_{1\gamma} Z_{2\delta} \nonumber\\
&&\cdot \delta^4(p-l_1-l_2)\nonumber\\
&\equiv&
-i 2\pi \delta^4(p-l_1-l_2)\cdot \mathcal{M}\quad,
\end{eqnarray}
where $\mathcal{M}$ denotes the Feynman amplitude.
The polarization vectors are perpendicular to the corresponding momentum

\begin{equation}
k_{1\mu} g_1^\mu=0\quad,\quad k_{2\mu} g_2^\mu=0\quad,\quad l_{1\mu} Z_1^\mu=0\quad,\quad l_{2\mu} Z_2^\mu=0.
\end{equation}
Further the calculation shall be considered in the center of mass system meaning that the following relations are valid

\begin{equation}
\vec k_1=-\vec k_2\quad,\quad \vec p=0\quad,\quad \vec l_1=-\vec l_2\quad.
\end{equation}
Thus, by using (\ref{Polarisationstensorgraviton}) and (\ref{Wexpression}) one obtains for the Feynman amplitude 

\begin{eqnarray}
\mathcal{M}&=&4A \cdot g_{1\alpha} g_{2\beta} \left[-\frac{1}{2}\eta^{\mu\nu}(k_1 \cdot k_2)(g_1 \cdot g_2) \right.\nonumber\\
&&\left.+(g_1 \cdot g_2)(k_1^\mu k_2^\nu)+(k_1 \cdot k_2)g_1^\mu g_2^\nu\right]\nonumber\\
&&\cdot \left[\frac{1}{2}(\eta_{\mu\rho}\eta_{\nu\sigma}+\eta_{\mu\sigma}\eta_{\nu\rho})-\frac{1}{3}\eta_{\mu\nu}\eta_{\rho\sigma}\right.\nonumber\\
&&\left.-\frac{1}{2m^2}(\eta_{\mu\rho}p_{\nu}p_{\sigma}+\eta_{\nu\sigma}p_{\mu}p_{\rho}+\eta_{\mu\sigma}p_{\nu}p_{\rho}+\eta_{\nu\rho}p_{\mu}p_{\sigma})\right.\nonumber\\
&&\left.+\frac{1}{3m^2}\eta_{\rho\sigma}p_\mu p_\nu+\frac{1}{3m^2}\eta_{\mu\nu}p_\rho p_\sigma+\frac{2}{3m^4} p_\mu p_\nu p_\rho p_\sigma\right]\nonumber\\
&&\cdot\left[-\frac{1}{2}\eta^{\rho\sigma}(l_1 \cdot l_2)(Z_1 \cdot Z_2)+(Z_1 \cdot Z_2)(l_1^\rho l_2^\sigma)\right.\nonumber\\
&&\left.+(l_1 \cdot l_2) Z_1^\rho Z_2^\sigma \right]\quad,
\label{Feynmanamplitude}
\end{eqnarray}
with $A=\frac{1}{(2\pi)^3 \sqrt{2k_{10}} \sqrt{2k_{20}} \sqrt{2l_{10}} \sqrt{2l_{20}}}\sum_n \frac{1}{\bar M_P^2} \frac{1}{p^2-m_n^2}$ and \\
$k_1 \cdot k_2=k_{1\mu}k_2^\mu$. Note, that the polarisation tensor of the graviton is on-shell meaning that $p^2=m^2$.
In general the differential cross section is related to the
squared Feynman amplitude $|\mathcal{M}|^2$ by the following expression
\begin{equation}
d \sigma=\frac{(2 \pi)^4}{2} \sum_\sigma |\mathcal{M}|^2 E \sqrt{E^2-m_Z^2}{\rm sin(\theta)} d \Omega\quad.
\end{equation}
Thus, the expression (\ref{Feynmanamplitude}) has to be squared. The summation over the polarization vectors of the Z-bosons 
and the gluons can be performed by using the relations 
\begin{equation}
\sum_\sigma g_\mu g_\nu^*=-\eta_{\mu\nu} \quad,\quad \sum_\sigma Z_\mu Z_\nu^*=\left(-\eta_{\mu\nu}+\frac{l_\mu l_\nu}{m_Z^2}\right)\quad.
\end{equation}
In order to obtain the total cross section one finally needs to integrate over the scattering angle $\theta$.
The infinite sum over Kaluza-Klein excitations $(\sum_n 1/(s-m_n^2))$ is treated with
dimensional regularization \cite{Giudice:1998ck}.
By taking the lowest dimensional contribution ($c_1=1$ and $c_i=0$ for $i \neq 1$) and
by choosing the regularization scale to be on the order of the new fundamental 
mass scale $\Lambda=M_D$ one finds
\begin{equation}
\sum_n \frac{1}{s-m_n^2}\approx\frac{\bar M_P^2\pi^{\frac{d}{2}}}{\Gamma (\frac{d}{2})M_D^4} \quad.
\label{eq_KKsum}
\end{equation}
Other approximations for the sum over the Kaluza Klein excitations 
are given in \cite{Han:1998sg,Hewett:2007st,Litim:2007iu}.
However, in our consideration we use equation 
(\ref{eq_KKsum}).\\
This leads to the cross section
\begin{equation}
\label{eq_ADDc}
\sigma(gg \rightarrow ZZ)=\frac{\pi^d \sqrt{\frac{\hat s}{4}}\sqrt{\frac{\hat s}{4}-m_Z^2}Z}{\Gamma^2 (\frac{d}{2}) M_D^8 30 \pi m_Z^4} 
\end{equation}
with
\begin{eqnarray}
Z&=&13.875 \hat s^4-115.625 \hat s^3 m_Z^2+311.0625 \hat s^2 m_Z^4\nonumber\\
&&-314.250 \hat s m_Z^6+98 m_Z^8\quad.
\end{eqnarray}
This is the inclusive cross section as a function of the number of extra dimensions $d$ and the
Planck mass $M_D$ in $4+d$ dimensions.

\section{Results}

To obtain the cross section for the (anti-)proton-proton process one finally integrates
over the parton distribution functions denoted by $f$ 
\begin{eqnarray}
\sigma(p(K_1) p(K_2)\rightarrow Z(l_1)Z(l_2))\ \ \ \ \ \ \ \ \ \ \ \ \ \ \ \ \ \nonumber\\
=\sum_{i,j}\int_0^1 dx_2 \int_0^1 dx_1 f_i(x_1,Q) f_j(x_2,Q)\ \ \ \ \ \ \ \ \ \  \nonumber\\
\sigma (\tilde p_i(k_1) \tilde p_j(k_2)\rightarrow Z(l_1)Z(l_2))\quad,
\end{eqnarray}
where $\tilde p_i$ denotes the parton (whether this is a quark or a gluon depends on
the index $i$) that is contributing to the process, $k_1=x_1 K_1$, and $k_2 = x_2 K_2$. 
The parton distribution functions (given in \cite{CTEQ6}) are evaluated 
at a scale of $Q=\sqrt{\hat s}$ for the process corresponding to figure (\ref{F1}) 
and at a scale of $Q=m_Z$ for the processes corresponding to figure (\ref{F2}), 
figure (\ref{F3}) and figure (\ref{F4}).  

\begin{figure}[ht]
\centering
\epsfig{figure=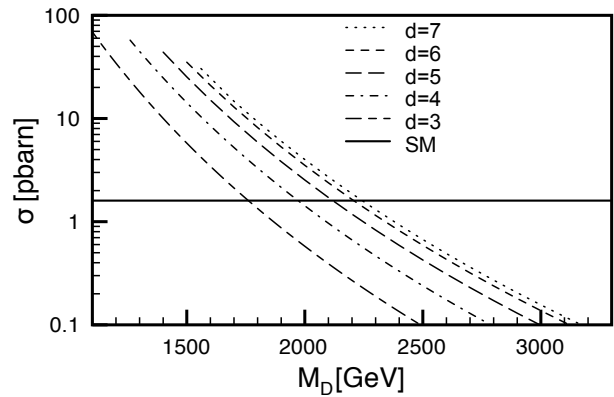,width=8.5cm}
\caption{\label{sig2}
Comparison between the inclusive $p\bar{p}\rightarrow ZZ$ production cross section
for the SM \cite{Campbell:1999ah}, the Fermilab \cite{Acosta:2005pq} 
limit and the ADD model ($d=3,4,5,6,7$) for $\sqrt{s}=2000$~GeV
in dependence of $M_D$.}
\end{figure}

In figure (\ref{sig2}) the inclusive cross section at the Fermilab energy of $2000$~GeV
is depicted as a function of the fundamental Mass scale $M_D$,
(starting from the lowest bound allowed by the unitarity constraint 
\cite{Giudice:1998ck}).
One sees that at $M_D > 2500$~GeV graviton mediation according to the theory decribed above does not have 
any observable influence on the ZZ rate at Fermilab \cite{Acosta:2005pq}.

\begin{figure}[ht]
\centering
\epsfig{figure=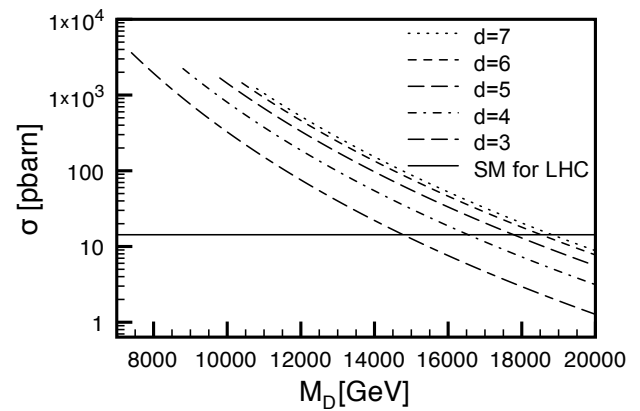,width=8.5cm}
\caption{\label{sig14}
Comparison between the inclusive $pp\rightarrow ZZ$ production cross section
for the SM \cite{Ohnemus:1995gb}  and the ADD model ($d=3,4,5,6,7$) for $\sqrt{s}=14000$~GeV
in dependence of $M_D$.}
\end{figure}

\begin{figure}[ht]
\centering
\epsfig{figure=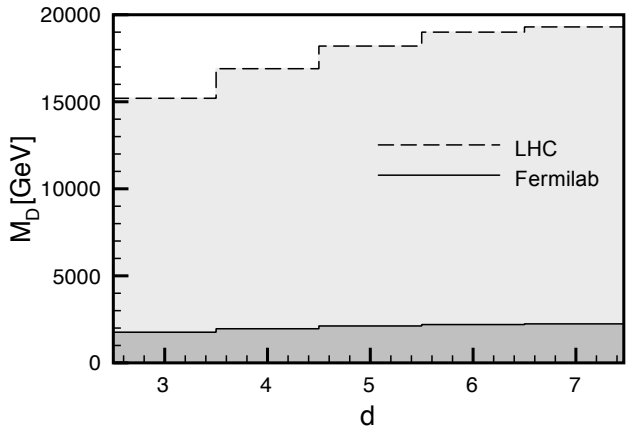,width=8.5cm}
\caption{\label{Mdschnitt}
ADD model parameter space accessible (i.e. when the additional cross section is at least as big as the SM cross section) in the ZZ-channel 
for the Tevatron and the LHC.}
\end{figure}

The same analysis is shown in figure (\ref{sig14}) for pp reaction at an energy of $14000$~GeV being available at LHC.
Here, the drastic difference between the SM di-Z-boson rate and the graviton mediated di-Z-boson rate might allow to observe effects of large extra dimensions 
even for a fundamental scale of $M_D\sim 18000$~GeV.
Figure (\ref{sig14}) shows that at $\sqrt{s}$=14000 GeV, the
ADD result dominates the SM prediction for small $M_D\ll \sqrt{s}$. 
This might reflect the fact,
that the regularization method and the perturbative field theory 
approach loose validity in this regime.
Therefore, it is more reasonable to take the results depicted in figures (\ref{sig2}) and (\ref{sig14})
only close to the regime of its validity $\sqrt{s}\sim M_D$.
This allows to state from which $M_D$ on experimental deviations from the SM ZZ rate at LHC \cite{Ohnemus:1995gb} should be expected.
In figure (\ref{Mdschnitt}) the testable parameter space is depicted for the LHC and Tevatron experiments
\footnote{It should be mentioned that by taking the approximation for the summation over the Kaluza-Klein
states given in \cite{Han:1998sg}, the $d$ dependence of the results differs decisively from the $d$ dependence obtained here}.
For experimental observation one would have to look for two high energetic and correlated lepton pairs in the final state as $Z\rightarrow l^+ l^-$.
By multiplying the total cross section with the branching ratio $\eta$ this cross section can be estimated.
The branching ratio can be obtained by taking the ratio of the couplings in the leptonic channels to
the couplings in all fermionic channels (the square appears because both Z-bosons should convert to a di-lepton pair).
\begin{equation}
\eta=\left(\frac{(\frac{1}{2}-\sin^2(\theta_W))^2+\sin^4(\theta_W)}{2-4\sin^2(\theta_W)+\frac{16}{3}\sin^4(\theta_W)}\right)^2
\approx 0.01 \quad,
\end{equation}
where $\theta_W$ is the Weinberg angle at the Z-scale and $\sin^2(\theta_W)\approx0.23$.

\section{Summary and Discussion}

Within the ADD model we have calculated the additional contribution 
to the double Z-boson production cross section in $pp,\bar pp$ reactions at high energies. 
The calculation is done to lowest order in $\sqrt{\alpha_{ew,strong}}$ and the ratio $m_{X}/M_D$. 
It is found that the standard ZZ-production rate would be significantly enhanced compared to the SM rate, 
if the ADD scale would be lower than $15000$~GeV in case of the LHC, respectively $1700$~GeV in case of the Tevatron. For the case of seven extra dimensions even 
$M_D=18000$~GeV could be tested at LHC.

In consideration of such results we
want to remind the reader that the magnitude of the ADD contribution (\ref{eq_ADDc}) directly
depends on the chosen regularization scale $\Lambda$ in equation (\ref{eq_KKsum}).
But a choice of $\Lambda=M_D$ is natural and because of
the fact that we only took the lowest dimensional term of 
the regularized graviton sum (see \cite{Giudice:1998ck}).

Therefore, if an enhancement above the SM ZZ-production rate would be observed at LHC energies,
it can provide important insights into the possibly higher dimensional structure of space-time.
\pagebreak\\
$Acknowledgement$:\\ We want to thank Elliot Lipeles for helpful comments . This work was supported by GSI, BMBF and FIAS.

\end{document}